# Design and Implementation Approach for Error Free Clinical Data Repository for the Medical Practitioners

[1]Prof. Kisor Ray, [2]Dr. Santanu Ghosh, [3]Dr.Mridul Das, [4]Dr.Bhaswati Ray

[1]*Techno India Agartala, Tripura University, Tripura, India*
[2,4]*Tripura Medical College, Tripura University, Tripura, India*
[3]*IGM Hospital, Govt. of Tripura*

*Abstract—* The modern treatment of any disease is heavily dependent on the medical diagnosis. Clinical data obtained through the diagnostics tests need to be collected and entered into the computer database in order to make a clinical data repository. In most of the cases, manual entry is an absolute necessity. However, manual entry can cause errors also, leading to wrong diagnosis. This paper explains how data could be entered free of error to reduce the chances of wrong diagnosis by designing and implementation of a simple database driven application.

## I. INTRODUCTION

A Clinical Data Repository (CDR) or Clinical Data Warehouse (CDW) is a real time database that consolidates data from a variety of clinical sources to present a unified view of a single patient [1]. Clinical data for the purpose of diagnosis is very important. The current treatment guidelines are mostly based on evidence [2]. A wrong data can cause a wrong diagnosis. Data obtained from the different tests require manual inputs into the computer database before generating the report for the patient. A good application with proper design can prevent entry of the wrong data as well as take care of hundreds of data fields through a simple program module with repetitive use of the module for all the fields. This kind of design and implementation is very important and useful for data entry purpose since large number of different types of field get covered and validated through a simple program module..

## II. MOTIVATION

Developed countries are having the infrastructure for the healthcare purpose. It's a multi-billion dollar industry in the organised sectors with pre-defined standards in practice. However, the developing countries including India lack those standards and practices barring a few exceptions in major cities. Most of the diagnostics labs across the country lack the minimum modern infrastructure and standards. Availability of cheaper personal desktops and laptops with the use of MS Office much of the reports are produced without the use of any defined standardized repository. The authors have mentioned about CDR in this paper starting from the title itself though they understand that CDR is quite a big thing from the point of implementation which requires including but not limited to initiative, funds, technology, knowledge, concept as well as awareness. This paper does not suggest to implement any centralized as well as standardized CDR rather it intends to provide guidelines for the health practitioners who use home grown systems for their patients. While developing smaller applications, practitioners may follow the simple design and implementation of the modules suggested through this paper which can significantly minimize the errors in the clinical data entry which eventually help them to build their own repository of patient database with clinical data relatively free of common as well as critical errors.

## III. THE PROBLEM

As mentioned earlier, in absence of any large scale standardized initiative, most of our health practitioners in India including those who are in the urban areas not much into the use of technology for capturing, storing and retrieving patient data as well as generation of reports including laboratory findings. A small numbers do use technology mainly for printing reports using MS office or similar software. So, in absence of database based application that too without validation, the use of technology is just for the cosmetic purpose. Even those who use home grown software lack the large scale alerts, verification & validations to identify the manual errors. These errors are introduced due to the mistakes made at the user levels which complicates the situation rather solving the problems of the patients.

IV. THE SOLUTION

What we have mentioned earlier is that we do not intend to produce a large centralized CDR rather we are focused to provide guidelines to standardize the smaller home grown initiatives at individual practitioners' level including the cosmetic uses. Should all these users use some kind of databases including those who are not using any, may adopt to build small applications without much effort. All they need is to use some databases which may include but not limited to Dbase/Xbase, FoxPro,MS Access, MySQL, PostGerySql to any other RDBMS like Oracle,SQLServer etc.., a form in the frontend and some reporting facilities with print option. The First step is to capture the patient information along with an Unique ID which could be used to identify the patient uniquely through search option, patient age , date of birth and contact details. The patient information form should have high degree of validation for important information like Unique ID, Date of birth and Age so that  no wrong information at least for these fields should not get entered into the patient repository. Next step is to capture the clinical information. This is a very important stage where n-numbers of fields may need rock solid validation and/or range indicator based on the subject to be entered. So far what we have mentioned should not be anything new but just a normal flow of any application. However, while we deal with the clinical data, we need to manage large number of fields at the frontend while reading and writing them to and from the backend repository. Providing range/validation/alert etc. at individual field level becomes a big task in terms of effort involved. Considering our intent to involve individual practitioners to develop some repositories through their personal initiatives, such higher level of efforts may obviously demotivate them. So, we need to design one or  a few very generic modules which could be used for all the fields very easily without much rewrite/modification but repeatedly thus minimizing the effort of development  of a minimal system which takes care of the usual errors including the standard and critical ones. What we know from the medical science is that the known clinical data  are mostly having certain range .So, while an operator enters these data, we can make him/her aware of the ranges at different levels so that she/he can  ill effort to make the mistakes. Even then, if any mistake is made, that should be under the notice of the supervisor through alerts and unless a supervised overwrite option is exercised for such alerts to accept the values beyond the ranges for any good reason, these data outside the normal ranges do not get into the repository, ensuring that no wrong data gets reported from the final outcome.

V. DESIGN PHILOSOPHY

The design philosophy is guided by the identification of the clinical data needed to be entered into the repository for the patients. However, while a transaction table in the backend repository deals with the patient data , a master table for the transactional fields are created with a known validated ranges with input from the specialists physicians. Creation of this kind of master tables and get them verified by the medical specialists is a very important process. So, when clinical data is entered into the transactional table of the backend repository for the visiting patient, field ranges are called from the master table and displayed on the screen at the run time to make the operator aware of the data as well as compare to what  she/he is entering. This could be implemented by just developing & using a simple module or subroutine and calling the same repeatedly by passing a parameter related to the transactional field dealt by the operator at that particular time. The design also ensures that should any real or apparent mistake even after the run-time comparison manages to slip and carries forward; later on, appears distinctly on the report itself before the supervisor signs off. Thus, providing a level 2 check up before the final report is delivered.

VI. IMPLEMENTATION

We know that a repository for a patient may include clinical laboratory test results, patient demographics, pharmacy information, radiology reports and images, pathology reports etc. For the implementation purpose let us take data on the laboratory test results as an example. The master table Tbl_M_ValueRange contains numerous data related to the blood test though a few are shown in the figure 1.0 below.

calling a simple module named as 'NormalValueRange'.

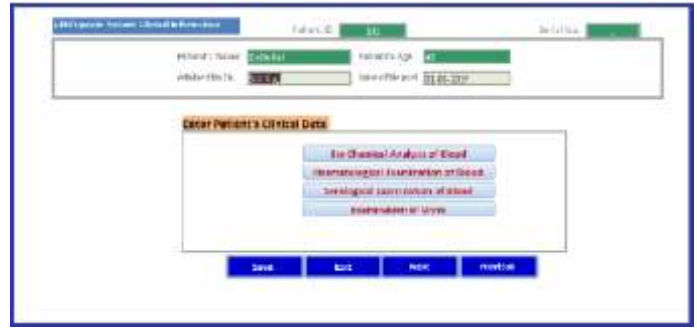

Fig 2.0 Data Entry Options for Tests

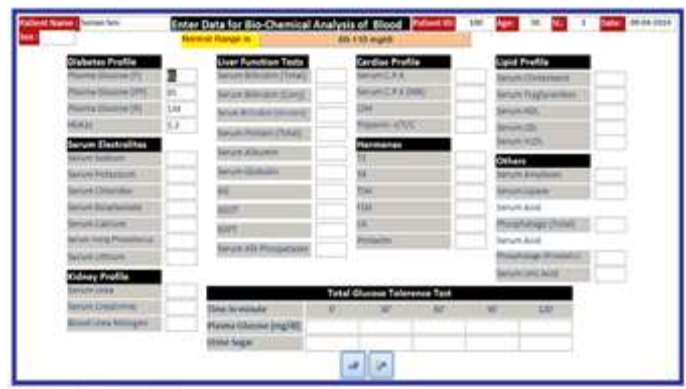

Fig 3.0 Frontend Data Entry with Range Values

This master table has three columns (fields) Field1->SLNO (Numeric Type) , Field2-> Test_Name (Text Type) and Field3-> ValueRange (Text Type). So, what we see that the blood related tests names are stored under the field2 along with its value ranges under the normal circumstances. The inputs into this kind of master table needs verification and validation by the specialist physicians since these will act to generate alerts and range violation warnings during a run-time transactional activities or we can say these are the baseline values for the purpose of comparisons. When input will be entered into the frontend form (Fig 3.0) , during the time of entry , a range for the value entered will be on the highlighted display by

' /*** Generic Module for Values with Range ***

Public Sub NormalValueRange (MyRangeSl As Variant)
 '**This is a public sub to get the normal value range ***

   '***** Open the ValueRange DB

    Set db = CurrentDb
    Set rst = db.OpenRecordset("Select Value_range from Tbl_M_ValueRange where [SLNO] = " & MyRangeSl & ";")
   If rst.EOF <> True Then
           Me.TxtNRange = rst("Value_range")
   Else
      MsgBox " Problem in opening the Normal Value Range Database Table!"
   End If

```
    rst.Close
    db.Close
End Sub
```

From the master table Tbl_M_ValueRange ( Fig 1 ) we see that the 1st record has a  column  test_name (Field2), a test name value like PlasmaGlucoseF (Field2) corresponds to the SLNO(Field1) with range values 60-110 mg/dl (Field3). While data will be entered into this field for a patient , the common module NormalValueRange will be called just by passing the value of the filed1(SLNO) which is equal to 1 in this case. Thus, we can write :

```
'******Invoke the Text Entry Field ******
Private Sub TxtPLGF_Enter()
Me.TxtNRange = ""
Me.TxtNRange.Visible = True
' **** Call the Sub and pass the parameter ****
NormalValueRange (1)
End Sub
```

As a result, during the transactional entry while data will be entered into the database table, the operator will get the highlighted display of the normal value range for that particular event ( in this example, value range of the plasma glucose F)  for which (SLNO=1) value is passed as parameter.

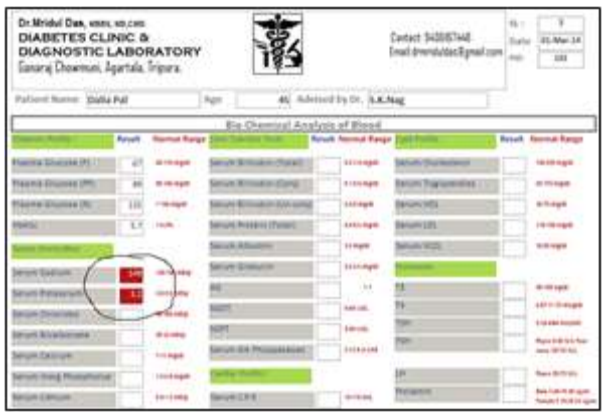

Fig 4.0  Report with alerts for range violation

At the end of the data entry for clinical data (for this particular example lab test results related to Bio-Chemical Analysis of Blood), report is prepared by pulling the data from the backend database transactional table for the patient. Should there be any range violation for any reason including but not limited to wrong data entry, the same comes up on the report for both the UL(Upper Limit) and LL(Lower Limit) violation distinctly for the supervisor (Fig 4.0 ) , thus enabling him/her to decide whether to accept or reject the data and/or inquire/recheck the entered data before finalizing as well as signing off the report for the patient. Thus, we can ensure two level checks by applying simple database techniques by reparative use of  a special module at the frontend and the report level.

## VII. CONCLUSION

In our introduction, we have talked about CDR what is frequently used by most of the healthcare providers specially to remain complaint with the local laws in the developed countries. We wish to have such system in place for us too. May be in the future such measures would be taken collectively by all the concerned related to healthcare. All we want to establish in our paper is that even in absence of any regulation and/or centralized initiative, our health practitioners can develop and use smaller low cost applications very easily without any significant effort which can eliminate much of the errors; if not all of them, following simple design guidelines and implementation of widely available databases which may or may not need any licensing cost depending upon the platform used. All we need to have such system at individual practitioners' level is the awareness as well as willingness to implement them with good collaboration between the physicians and the IT professionals.

## REFERENCES

[1] MacKenzie, S. L.; Wyatt, M. C.; Schuff, R.; Tenenbaum, J. D.; Anderson, N. (2012). "Practices and perspectives on building integrated data repositories: Results from a 2010 CTSA survey". Journal of the American Medical Informatics Association 19 (e1): e119–e124. doi:10.1136/amiajnl-2011-000508. PMC 3392848. PMID 22437072

[2] Deo S. Computerized clinical database development in oncology. Indian J Palliat Care 2011;17, Suppl S1:2-3